\documentclass[aps,prb,reprint,superscriptaddress,amssymb,amsmath]{revtex4-1}
\usepackage{graphicx}
\usepackage{dcolumn}
\usepackage{bm, color}
\usepackage{bbm}
\usepackage{braket}
\usepackage{tabularx}

\begin{document}

\title{Emerging chiral edge states from the confinement of a magnetic Weyl semimetal in Co$_3$Sn$_2$S$_2$}

\author{Lukas Muechler}
\affiliation{Center for Computational Quantum Physics, The Flatiron Institute, New York, New York, 10010, USA}
\author{Enke Liu}
\affiliation{Max Planck Institute for Chemical Physics of Solids, D-01187 Dresden, Germany}
\affiliation{Institute of Physics, Chinese Academy of Sciences, Beijing 100190, China}
\author{Jacob Gayles}
\affiliation{Max Planck Institute for Chemical Physics of Solids, D-01187 Dresden, Germany}
\author{Qiunan Xu}
\affiliation{Max Planck Institute for Chemical Physics of Solids, D-01187 Dresden, Germany}
\author{Claudia Felser}
\email{Claudia.Felser@cpfs.mpg.de}
\affiliation{Max Planck Institute for Chemical Physics of Solids, D-01187 Dresden, Germany}
\author{Yan Sun}
\email{ysun@cpfs.mpg.de}
\affiliation{Max Planck Institute for Chemical Physics of Solids, D-01187 Dresden, Germany}

\date{\today}
\begin{abstract}

The quantum anomalous Hall effect (QAHE) and magnetic Weyl 
semimetals (WSMs) are topological states induced by intrinsic magnetic moments and spin-orbit coupling. Their similarity suggests the possibility of achieving the QAHE by dimensional confinement of a magnetic WSM along one direction. 
In this study, we investigate the emergence of the QAHE in the two-dimensional (2D) limit of magnetic WSMs due to finite size effects in thin films and step-edges. We demonstrate the feasibility of this approach with effective models and real materials. To this end, we have chosen the layered magnetic WSM Co$_3$Sn$_2$S$_2$, which features a large anomalous Hall conductivity and anomalous Hall angle in its 3D bulk, as our material candidate. In the 2D limit of Co$_3$Sn$_2$S$_2$ two QAHE states exist depending on the stoichiometry of the 2D layer. One is a semimetal with a Chern number of 6, and the other is an insulator with a Chern number of 3. The latter has a band gap of 0.05 eV, which is much larger than that in 
magnetically doped topological insulators.
Our findings naturally explain the existence of chiral states in step edges of bulk Co$_3$Sn$_2$S$_2$ which habe been reported in a recent experiment at $T = 4K$ and present a realistic avenue to realize QAH states in thin films of magnetic WSMs.
\end{abstract}

\maketitle

\section{Introduction}

The quantum anomalous Hall effect (QAHE) is the quantized 
version of the anomalous Hall effect (AHE). In the QAHE, the 
transverse anomalous Hall conductance (AHC) is quantized 
in units of $e^2/h$ and the longitudinal resistance is reduced to zero
~\cite{Hall1879, Hall1881,Nagaosa2010, Haldane1988}. In contrast to the ordinary
AHE, the QAHE is topologically protected, and is characterized by a non-zero
Chern number $n$ and chiral edge states in the bulk band gap. Contrary to the
quantum Hall effect, which features the formation of Landau levels induced by an external 
magnetic field, the QAHE originates
from intrinsic magnetic moments and spin orbit coupling (SOC).
Because of the absence of back scattering, the electron currents carried by the
edge states cannot be dissipated~\cite{Halperin1982}. Therefore, 
realization of the QAHE and the utilization of
its chiral edge state may lead to new low energy consumption quantum electronic devices.
It was further proposed that interfaces of QAHE states and superconductors  can
generate chiral Majarona Fermions (MFs)~\cite{Xiaoliang_2010_QHS} via the proximity 
effect. These MFs are thought to be a fundamental ingredient for the development of 
topological quantum computing.

The QAHE was first proposed in 1988 by F. D. M. Haldane 
via a theoretical model defined on a honeycomb 
lattice~\cite{Haldane1988}. However, because of the special requirements of this model,
the realization of the QAHE in real materials took more than
twenty years to achieve. Thanks to developments in topological band theory and thin film growth
techniques, the effect was observed in counterparts of the QAHE, the quantum spin 
Hall effect (QSHE) and topological insulators (TIs). The first QAHE was observed in 
chromium-doped (Bi,Sb)$_2$Te$_3$, where the TI gains magnetic order through Van 
Vleck paramagnetism~\cite{Yu2010, Chang2013}. 
Based on the chiral edge states originating from Cr-doped (Bi,Sb)$_2$Te$_3$, chiral
MFs were recently observed in the heterostructure of QAHE states and superconductors,
which created strong enthusiasm for the comprehensive study of the QAHE and MFs using both theoretical and
experimental approaches~\cite{He2017,liu2016quantum,weng2015quantum,ren2016topological,wang2015quantum}. 

Besides magnetically doped topological insulators~\cite{liu2008quantum,Yu2010, Chang2013},
much effort was also devoted to the search for new QAHE materials 
based on different mechanisms~\cite{liu2016quantum,weng2015quantum,ren2016topological,wang2015quantum},
such as magnetic and antiferromagnetic semiconductors~\cite{liang2013,he2017qah,Sheng2017},
heterostructures of topological insulators (TIs) and ferromagnetic insulators
~\cite{Eremeev2013,Eremeev2015,Ferhat2016,Otrokov2017,Otrokov2017jpte,Joseph2017}
heterostructures of heavy atom layers and magnets~\cite{Garrity2013, Garrity2014, Liu2015,gangxu2015,Si2017},
honeycomb lattice (like graphen, silicene, and transition metal oxide with perovskite
structure and pyrochlore oxide  along (111) direction) 
with proximity coupling to ferromagnetic and antiferromagnetic
insulators~\cite{qiao2010quantum,nandkishore2010quantum,tse2011quantum,
zhang2011spontaneous,ezawa2012valley,ezawa2013photoinduced,zhang2015,
zhang2015prb,qiao2014,xiao2011interface,ruegg2011topological,
hu2012topological,zyuzin2012weyl,liang2013electrically}, 
asl well as thin films of 3D magnetic toplogical states~\cite{xu2011chern},
and absorption of atoms on quantum spin Hall insulators~\cite{Shu-Chun2014PRL}, etc.
For example, a gap around 100 meV was observed by ARPES in the heterostructure of
MnBi$_2$Se$_4$/Bi$_2$Se$_3$~\cite{Hirahara2017}.

Recently, another topological metallic state, the Weyl semimetal (WSM), 
was experimentally determined~\cite{Xu2015TaAs,Lv2015TaAs}.
Magnetic WSM with layered structures can 
be understood as coupled 2D QAHE layers~\cite{Wan2011,Burkov:2011de}.
Motivated by this understanding and the development of WSMs,
we propose another method of efficiently realizing the QAHE state in a real material.
In WSMs, the conduction and valence bands linearly 
touch each other in 3D momentum space at Weyl points, 
which behave as monopoles of Berry curvatures 
with positive and negative chirality.
According to symmetries, WSMs can be classified into two groups, 
those with and without time reversal
symmetry. Since the Berry curvature is odd under time reversal, 
the magnetic WSMs often
host a strong AHE originating from the Weyl points. Since Weyl points can only 
be defined in 3D momentum space, they require translation symmetry in 3D. 
Therefore, the Weyl point can be removed by opening
a band gap through the breaking of periodic boundary conditions 
along one direction. If band ordering is maintained during this 
procedure, it is possible to obtain a QAHE phase in the 2D limit of 
the magnetic WSMs. On the other hand, since the QAHE exists only in the 
extreme 2D limit, magnetic WSMs with layered crystal structures are 
preferred material candidates. \\

\section{Outline}
The paper is organized as follows:
In Sec.~\ref{sec:model} we review the proposal to realize a QAH state in thin films of a magnetic WSM based on model calculations.
Using ab-initio calculations, we show in Sec.~\ref{sec:abinitio} that thin films of the magnetic WSM Co$_3$Sn$_2$S$_2$ are ideal candidates to realize the QAH state in two different stochiometries.
In the last section (Sec.~\ref{sec:stepedge}), we discuss how these findings explain the recent observation of chiral modes on step edges of bulk Co$_3$Sn$_2$S$_2$.

\section{Model discussion}
\label{sec:model}
To evaluate the feasibility of this proposal, we first analyzed the 2D
limit of the Weyl semimetal using a toy two band model,
$H=A(k_x\sigma(x)+k_y\sigma(y))+M(k)\sigma_{z}$, with $M(k)=M_0-M_1(k_x^2+k_y^2+k_z^2)$~\cite{Luhaizhou2015}.
As shown in Fig. 1(a), this model describes a magnetic WSM with one pair of Weyl
points located at $\pm k_{W}$
along the $k_z$-axis and the Fermi level (E$_F$) lies just at the Weyl point. Owing to the Berry
flux between this pair of Weyl points, the anomalous Hall conductivity reaches 
a peak value at E$_F$ with $\sigma_{AH}=(k_{w}/\pi)(e^{2}/h)$, which is proportional to the separation
of the Weyl points, see Fig. 1 (b) and (c)~\cite{Luhaizhou2015}. Breaking the periodic boundary
condition in the $z$ direction, the Weyl points are removed by the opening of the band gap.
Depending on the interaction strength between the top and bottom sides of the film,
the band order can change as the thickness varies,
resulting in a quantized jump of the AHC from zero
to a finite number. From the parameters specified in this study, the interactions between the top and bottom surfaces remove the band inversion of the 3D bulk state for thicknesses of one and two unit cells.
Upon increasing the film thickness above three unit cells, the quantum confinement effect
weakens and the band inversion remains, which leads to a
quantized AHC and chiral edge state in the band gap, as shown
in Fig. 1(d–f). Therefore, the mechanism for obtaining the QAHE from magnetic WSMs is principally
allowed, but its realization from available materials remains challenging.

\begin{figure}[htb]
\centering
\includegraphics[width=0.5\textwidth]{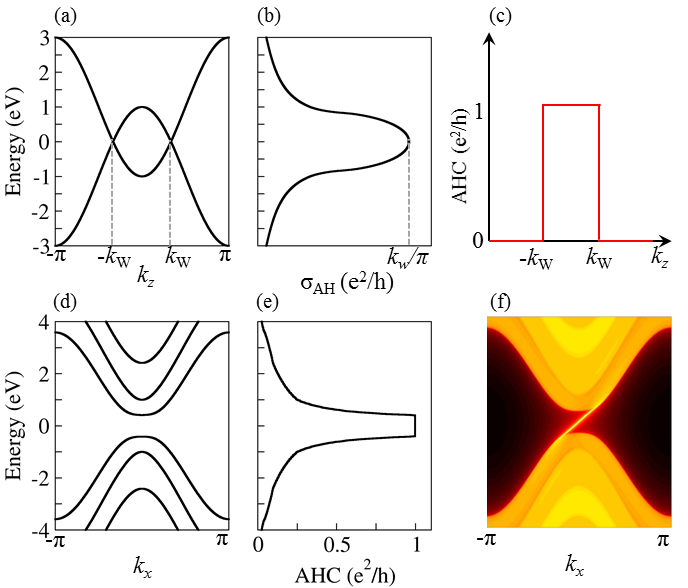}
   \caption{
(a) Energy dispersion along $k_z$ and (b) energy dependent anomalous Hall conductivity
in the WSM effective model. 
(c) A quantized 2D-anomalous Hall conductance (AHC) exists between
a pair of Weyl points for fixed $k_z$ between $-k_w < k_z < k_w $.
(d) Energy dispersion of the film from the WSM effective model with a thickness of three unit cells.
(e) The anomalous Hall conductance (AHC ) is quantized in the band gap of the film.
(f) Energy dispersion of the edge states of the film. We have projected the $k\cdot p$ model to a 
cubic lattice with a lattice constant of 1 \AA, and parameters of $A$=1.0 eV \AA, 
$M_0$=1 eV, and $M_1$=1 eV \AA$^2$}
\label{fig:model}
\end{figure}

To date, many theoretical magnetic WSMs have been proposed, some of which feature 
strong AHEs
~\cite{Wan2011, Xu2011,Wang_Heusler_2016,Chang2016,Kubler2016}.
Since the QAHE is a 2D insulator, WSMs
with large AHE, low charge carrier
density, and layered lattice structure are preferred. In this study,
we choose the experimentally well characterized
quasi-2D material Co$_3$Sn$_2$S$_2$ to examine the possibility of realizing
a QAHE based on a magnetic WSM. Co$_3$Sn$_2$S$_2$ is a half metal with a
rhombohedral lattice
structure in the space group $R\overline{3}m$ (No. 166)
~\cite{Weihrich2005, Vaqueiro2009, Schnelle2013}.
The magnetic Co atoms are arranged in a Kagome lattice in the $x-y$ plane 
with magnetic moments
along the $z$ axis. When the lattice is placed in a hexagonal setting, 
Co$_3$Sn forms a quasi-2D layer and a
sandwich between the S atoms. The quasi-2D layer Co$_3$SnS$_2$ are 
connected by another
Sn atom in the $z$ direction, as shown in Fig. 2(a).

\begin{figure}[htb]
\centering
\includegraphics[width=0.5\textwidth]{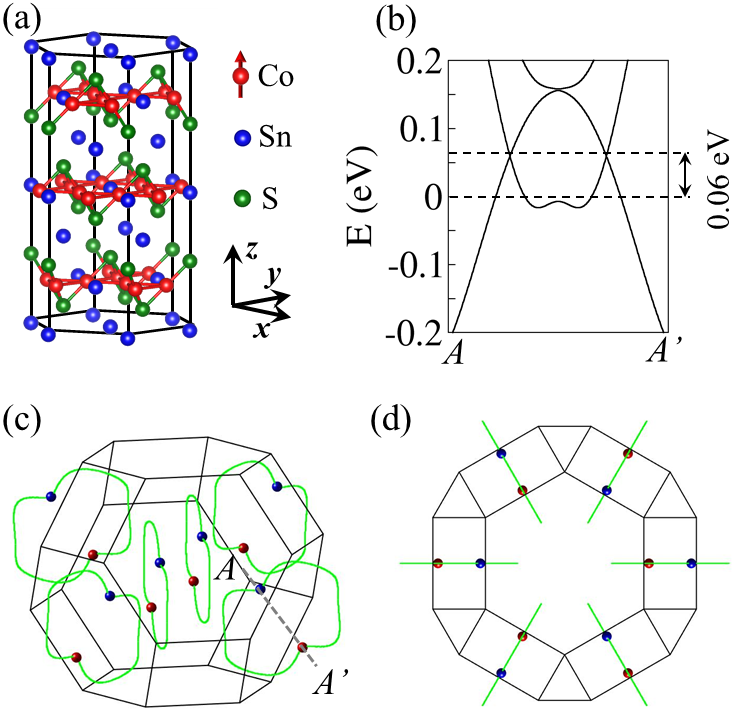}
   \caption{
(a) Lattice structure of Co$_3$Sn$_2$S$_2$. The spin polarization of the Co atom is along the
$z$ direction. (b) Energy dispersion of Co$_3$Sn$_2$S$_2$ along one pair of Weyl points with
opposite chirality. The labels for the $k$ points are given in (c).
(c, d) Side view and top view of the nodal lines and Weyl points distribution in BZ, respectively. 
The red and blue dots represent opposite chirality of the Weyl points.
}
\label{fig:bulk}
\end{figure}

\section{Single layer limit}
\label{sec:abinitio}
Recently, large anomalous Hall conductivities up to 1100 S/cm were observed 
in Co$_3$Sn$_2$S$_2$, originating from the SOC induced node-line-like band anticrossings and Weyl
points, see Fig. 2(c, d). Owing to the low charge carrier density, the anomalous
Hall angle reaches as high as 20\%, which has not been observed in other 
compounds~\cite{Enk_2017,Wangqi2017}.
In this material, the Weyl points are very close to E$_F$ and there is almost no 
interference from trivial bands, in contrast to other proposed magnetic WSMs.
These properties make Co$_3$Sn$_2$S$_2$ an ideal candidate to realize a QAH state with chiral edges modes in the 2D limit due to finite size effects.
Because of the weak bonding between Sn and Co$_3$SnS$_2$ layers, it should be possible to achieve
the 2D films and step states by breaking the weak bonds. Therefore, Co$_3$Sn$_2$S$_2$
provides an ideal platform to study the interplay between the QAHE and WSMs
experimentally.

\begin{figure}[htb]
\centering
\includegraphics[width=0.5\textwidth]{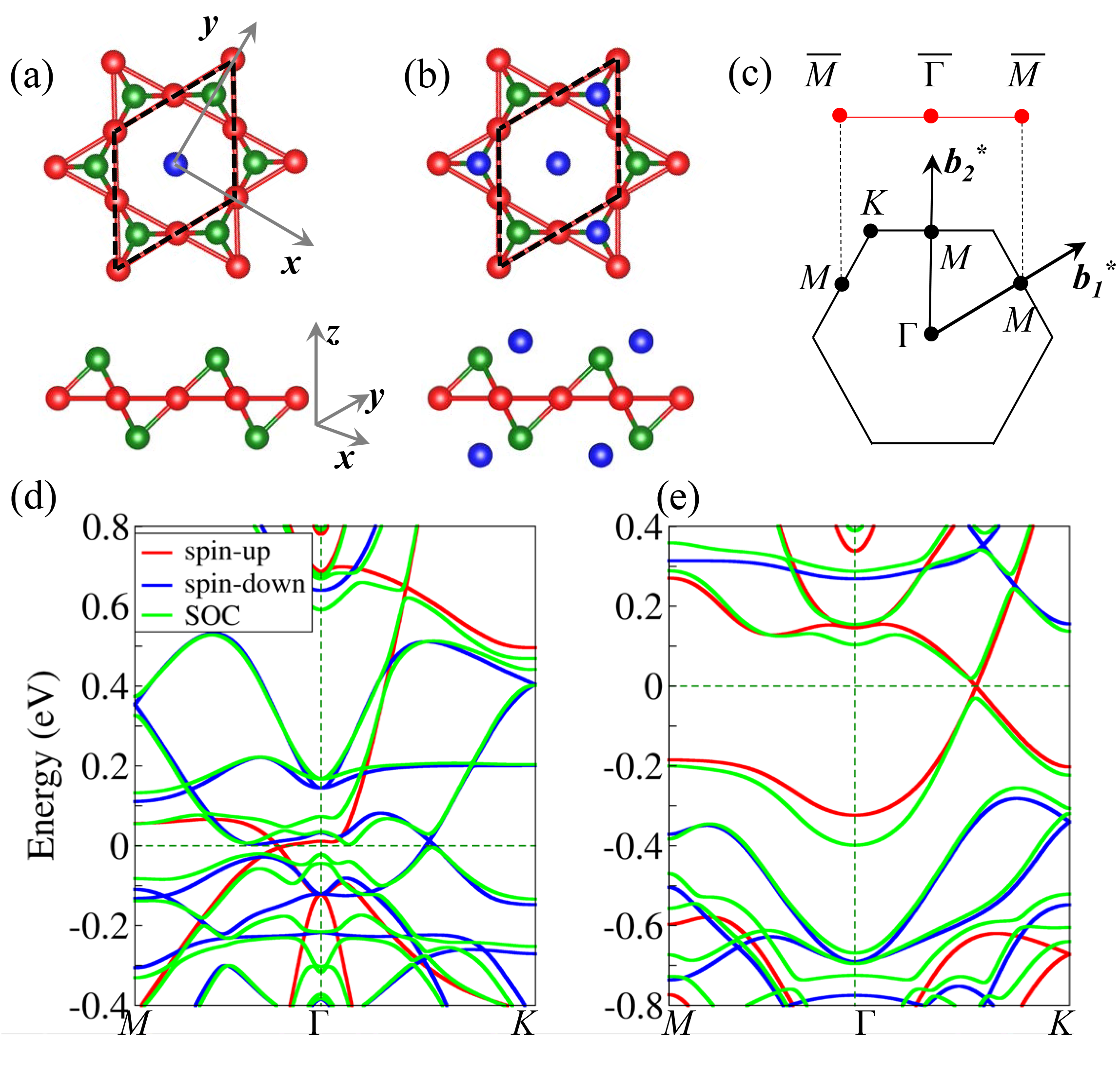}
   \caption{
Lattice structure of (a) Co$_3$SnS$_2$ and (b) Co$_3$Sn$_3$S$_2$ films from the top (upper panel)
and side (lower panel) view. (c) Corresponding 2D BZ and its 1D projection. The energy dispersion 
along high symmetry lines as shown for (d) Co$_3$SnS$_2$ and (e) Co$_3$Sn$_3$S$_2$ with and 
without the inclusion of SOC.
}
\label{fig:film}
\end{figure}

Two possible terminations exist for cleaved Co$_3$Sn$_2$S$_2$,
one with a Sn termination and the other with a S termination.
The Co atoms form a Kagome lattice, with the Sn atom in the center.
Sandwiching the Kagome lattice of Co$_3$Sn between two S layers, a minimized
thickness can be obtained for the S-terminated Co$_3$SnS$_2$ film, see Fig. 3(a).
By further sandwiching the Co$_3$SnS$_2$ with neighboring Sn layers, the other
minimum thickness can be obtained for the Sn-terminated Co$_3$Sn$_3$S$_2$ film.
To obtain the magnetic ground state, the total energies of magnetization
along the (001), (100), (010), and (110) directions were compared for both the bulk
state and the two possible structures for the monolayers via density functional
theory (DFT) calculations ~\cite{kresse1996, perdew1996}. For the bulk state,
the total energy in (001) phase is around 1.4 meV lower than that with magnetization
oriented in the $x-y$ plane, which is consistent with the experimental
measurements~\cite{Vaqueiro2009,Enk_2017,Wangqi2017}.
Similarly, the two monolayer structures also preferred a magnetic structure with
magnetization along the (001) direction, however the energy differences are only around
0.4 and 0.1 meV in Co$_3$Sn$_3$S$_2$ and Co$_3$SnS$_2$ films, respectively, much
lower than in the bulk with $T_c \simeq 200K$. Owing to the 2D magnetic structure and the small anisotropic energy,
the dipole-dipole iteraction might play an important role for the spin oriented in plane.
However, the dipole-dipole interaction is around one order of magnitude 
weaker than the anisotropic energy, due to the small moment of the Co ions.
see Fig. 4. Therefore, the magnetic are expected to along (001) direction. Based on the
 anisotropic  energy, the magnetic order temperature for Co$_3$Sn$_3$S$_2$ and 
Co$_3$SnS$_2$ freestanding monolayers are around 5.0 and 1.0 K, respectively. 
The anisotropy can increase in the presence of a substrate, which in turn can 
increase the ordering temperature.

\begin{figure}[htb]
\centering
\includegraphics[width=0.5\textwidth]{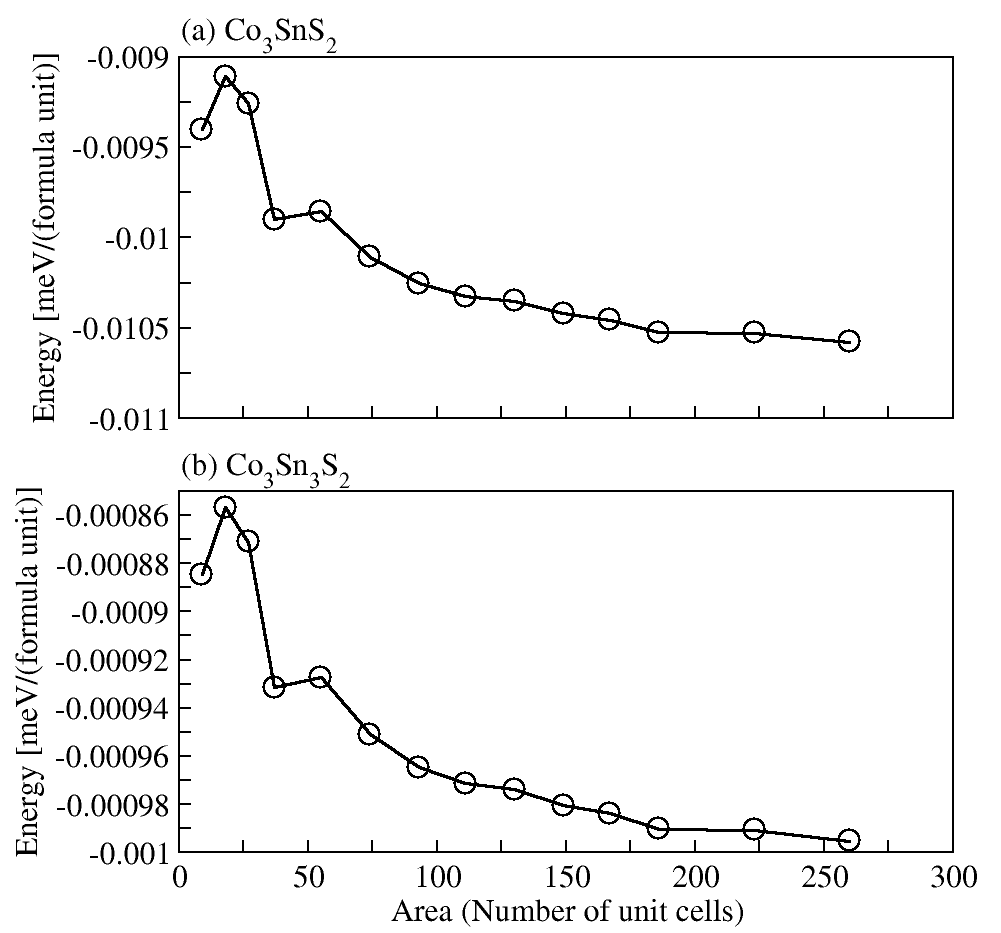}
   \caption{
Convergence of the dipole-dipole interaction along with the increasing 
of number of in-plane unit cells.
   }
\label{fig:dd}
\end{figure}

\begin{table*}[ht]
  \centering
  \caption{The comparison of total energies with magnetic polarization along (001) and other directions.
The energy differences are in the unit of of meV/(formula unit).
}
    \label{tab:table1}
  \begin{tabular} {cccccc}
  \hline
  \hline
                           & E$_{tot}$(100)-E$_{tot}$(001) & E$_{tot}$(010)-E$_{tot}$(001) &  E$_{tot}$(110)-E$_{tot}$(001)    \\
  \hline
  Co$_9$Sn$_6$S$_6$ (bulk) &  1.4                          &  1.4                          & 1.4                               \\
  Co$_3$Sn$_3$S$_2$        &  0.5                          &  0.4                          & 0.4                                \\
  Co$_3$SnS$_2$            &  0.2                          &  0.2                          & 0.1                                \\
\hline
\hline
\hline
  \end{tabular}
\end{table*}

Without SOC, the Co$_3$SnS$_2$ film shows a semi-metallic band structure.
In contrast to the half-metallic state in the 3D bulk, both spin-up and
spin-down channels appear close to the Fermi level in the Co$_3$SnS$_2$ 
films. Owing to the mirror planes of $M_x$ and $M_y$, both spin-up and 
spin-down channels form the linear band crossing around $E_F$, which 
exist on the high symmetry lines along $M-\Gamma$ and $\Gamma-K$, 
respectively, see Fig. 3(d). After SOC is considered, the SU(2) symmetry 
is broken, leading to gapping of the linear band crossings.

Viewing the energy dispersion in a larger energy window, it is clear that another
linear band exists crossing $\Gamma-K$ with its chemical potential shifting up 
by approximately 0.6 eV and 4 electrons, corresponding to the filling expected 
for films with Sn termination. This system should act as a semi-metal from the 
perspective of transport, by simply shifting the chemical potential to the upper 
linear band crossing point, see Fig. 3(d).
However, the Sn layers affect the filling and change the dispersion of the
bands around $E_F$. As shown in Fig. 3 (e), after further sandwiching by Sn layers, 
the system has an ideal semi-metallic state with the Fermi surface composed 
solely of the linear crossing points. Similar to Co$_3$SnS$_2$, this linear band 
crossing is broken by SOC with the opening of a band gap. Compared to Co$_3$SnS$_2$, 
the Sn terminated film has a global band gap at approximately 0.05 eV, which is much 
larger than that in magnetic-impurity-doped TIs.

\begin{figure}[htb]
\centering
\includegraphics[width=0.5\textwidth]{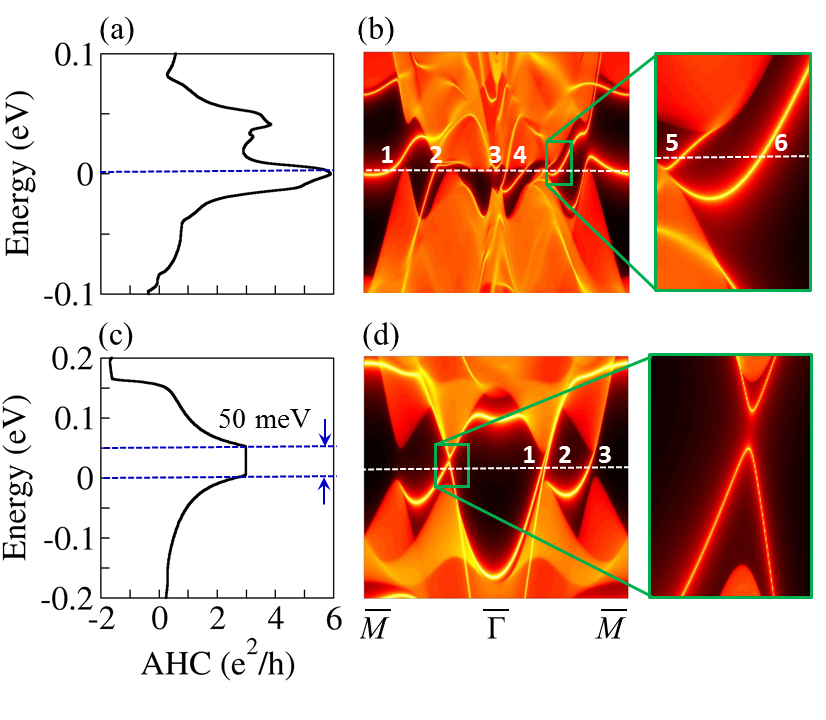}
   \caption{
(a, b) Energy dependent AHC for the Co$_3$SnS$_2$ film. The peak value with AHC=6 $e^2/h$ appears at a charge neutral point.
(b) Energy dispersion of the edge states for the Co$_3$SnS$_2$ film. The local energy dispersion around edge states No.5
and No.6 are given on the right.
(c) Quantized AHC (3 $e^2/h$) appears in the band gap for the Co$_3$Sn$_3$S$_2$ film.
(d) Edge states of Co$_3$Sn$_3$S$_2$ film. The trivial dangling bond states are zoomed in on the right side.
}
\label{fig:ahc}
\end{figure}

To confirm the quantization of the anomalous Hall conductance (AHC), 
we calculated the energy dependent
AHC by integrating the Berry curvature over the whole BZ.
The AHC for Co$_3$SnS$_2$ can reach up to almost 6 $e^2/h$ at the
charge neutral point. However, since Co$_3$SnS$_2$ does not have a global band gap, its
AHC changes sharply as the energy varies around $E_F$, and
the maximum value only appears at an energy point, see Fig. 5(a). 
Though Co$_3$Sn$_3$S$_2$ has four more electrons than Co$_3$SnS$_2$,
only three of them have non-zero chern number (--1), leading to the Chern number
jump from 6 to 3.
In contrast the the semimetallic state in Co$_3$SnS$_2$, the film with a 
stoichiometry of Co$_3$Sn$_3$S$_2$ has a constant quantized AHC in the energy range
from $E_F$ to $E_F$+50 meV, as shown in Fig. 5(c). Thus, both the Co$_3$SnS$_2$ and
Co$_3$Sn$_3$S$_2$ films host nontrivial topological electronic structures with 
non-zero Chern numbers of 6 and 3, respectively.

A typical feature of the quantum anomalous Hall insulators and semi-metals is their chiral
edge state. To calculate the edge states, we projected the Bloch wave functions into
Wannier orbitals ~\cite{Mostofi2008}, and constructed a tight binding model Hamiltonian.
The edge state was considered under open boundary conditions with
the half-infinite one-dimensional model using the iterative Greens function
method~\cite{Sancho1984,Sancho1985}. The energy dispersions of the edge states
for films with the nominal compositions Co$_3$SnS$_2$ and Co$_3$Sn$_3$S$_2$ are
shown in Fig. 5(b) and (d), respectively.
Due to the semi-metallic properties of the monolayers with the
stoichiometry Co$_3$SnS$_2$, the projections of the bulk conduction and
valence are separated by a gap, which are connected by six edge bands.
The edge bands all posses a positive Fermi velocity, corresponding to
a Chern number of 6. In the monolayers with the stoichiometry
Co$_3$Sn$_3$S$_2$, three topological protected channels exist
on the edge with positive Fermi velocity, which is consistent with
a Chern number of 3 obtained from the calculation of the quantized AHC,
see Fig. 5(d).  Since the properties of the edge states are strongly
dependent on the nature of the edge termination, the topological
edge-states can be distinguished from trivial dangling bond states,
since the dangling bonds are not topologically protected.
They do not connect the valence and conduction states of the bulk,
as can bee seen in the “zoom in” in Fig. 5(d), and should be easily
annihilated by chemical engineering of the edge termination.

\section{Chiral modes at step edges}
\label{sec:stepedge}
Although a QAHE is expected in monolayers of the stoichiometry Co$_3$SnS$_2$ and Co$_3$Sn$_3$S$_2$,
their free standing form is not expected to be stable without a substrate, indicated by phonon frequencies of the free standing slabs.
A suitable substrate could stabilize these monolayers at higher temperatures due to kinetic effects, however this is beyond the possibilities of DFT to predict and requires additional experimental input.
However, our results further imply the presence of chiral modes at step edges on bulk Co$_3$Sn$_2$S$_2$, which could be easily measured by STM experiments. 
A step edge can be modeled as a single QAH layer that is coupled to the bulk WSM. If the coupling between the QAH layer and the bulk can be treated perturbatively as expected in a layered material, the step edge is expected to host chiral edge states, similarly to the QSH step edge states observed in WTe$_2$~\cite{peng2017observation}. 
Even in the case of stronger coupling between the layers, the 'terrace construction' of \citet{Hsieh2016Bulk} suggests the presence of spatially separated, counterpropagating chiral states at step edges that could be observed in STM experiments, as the combined Chern number of the top layer and WSM has to vanish.
In this setup, one would expect to observe the edge states at temperatures close to the bulk $T_c \simeq 200K$, which could open a pathway to more detailed studies of topological chiral states at higher temperatures.
Indeed, such states have recently been reported in STM experiments on step-edges at $T = 4K$~\cite{howard2019observation} as a direct consequence of the topological non-trivial electronic structure in the thin film limit.
\\
\section{Summary}
In summary, we theoretically studied the QAHE from the quantum confinement
of the magnetic WSMs. 
Since the WSM can only be defined in 3D momentum space, a method for the phase
transition from WSM to QAHE is breaking the translational symmetry along one direction in the
WSM. Considering the 2D topological insulating state in the QAHE, the WSMs with strong AHE, 
low charge density, and layered lattice structure are preferred candidates. 
Considering these material requirements, we focused on the layered 
magnetic WSM Co$_3$Sn$_2$S$_2$ as a candidate for the realization of QAHE from WSMs.
In the 2D limit of Co$_3$Sn$_2$S$_2$, we find two possible QAHE states. 
One is a semi-metal with a Chern number of~6, and the other is an insulator with a 
Chern number of~3 and band gap of 0.05 eV. 
Our findings are supported by the recent observation of chiral modes at step edges of bulk Co$_3$Sn$_2$S$_2$, which follows from the non-zero Chern numbers in the 2D limit of Co$_3$Sn$_2$S$_2$~\cite{howard2019observation}.

\begin{acknowledgments}
This work was financially supported by the ERC Advanced Grant No. 
291472 `Idea Heusler', ERC Advanced Grant No. 742068--TOPMAT, and 
Deutsche Forschungsgemeinschaft DFG under SFB 1143. 
EL acknowledges support from the Alexander von Humboldt foundation of Germany for his
Fellowship and from National Natural Science Foundation of China for his Excellent Young
Scholarship (No. 51722106).
LM would like to thank the MPI CPFS where part of the work was performed.
and the authors thank Binghai Yan for helpful discussions.
The Flatiron Institute is a division of the Simons Foundation.
\end{acknowledgments}

\bibliography{TopMater}

\end{document}